\documentclass{ar-1col}
\usepackage[numbers]{natbib}
\usepackage{url}
\usepackage{amssymb}
\usepackage[normalem]{ulem}
\jname{\uline{When citing this paper, please use the following:} Liu Y, Ni K-K, 2022. Bimolecular chemistry inthe ultracold regime. Annu. Rev. Phys. Chem.: Submitted. DOI: 10.1146/annurev-physchem-090419-043244}

\jyear{2022}
\begin{document}

% Page header
\markboth{Liu, Ni}{Ultracold Molecular Chemistry}

% Title
\title{Bimolecular chemistry in the ultracold regime}

%Authors, affiliations address.
\author{Yu Liu,$^{1}$ Kang-Kuen Ni$^{2,3}$
\affil{$^1$Time and Frequency Division, National
Institute of Standards and Technology, Boulder, Colorado, 80305, USA; email: Yu.Liu-1@colorado.edu; ORCID: 0000-0002-5228-1997}
\affil{$^2$Department of Chemistry and Chemical Biology, Harvard University, Cambridge, Massachusetts, 02138, USA; email: ni@chemistry.harvard.edu; ORCID: 0000-0002-0537-0719}
\affil{$^3$Harvard-MIT Center for Ultracold Atoms, Cambridge, Massachusetts, 02138, USA}}

%Abstract
\begin{abstract}
Advances in atomic, molecular, and optical (AMO) physics techniques allowed the cooling of simple molecules down to the ultracold regime ($\lesssim$ 1 mK), and opened the opportunities to study chemical reactions with unprecedented levels of control. This review covers recent developments in studying bimolecular chemistry at ultralow temperatures. We begin with a brief overview of methods for producing, manipulating, and detecting ultracold molecules. We then survey experimental works that exploit the controllability of ultracold molecules to probe and modify their long-range interactions. Further combining the use of physical chemistry techniques, such as mass spectrometry and ion imaging,  significantly improved the detection of ultracold reactions and enabled explorations of their dynamics in the short-range. We discuss a series of studies on the reaction KRb + KRb $\rightarrow$ K$_2$ + Rb$_2$ initiated below 1~$\mu$K, including the direct observation of a long-lived complex, the demonstration of product rotational state control via conserved nuclear spins, and a test of the statistical model using the complete quantum state distribution of the products.
\end{abstract}

%Keywords, etc.
\maketitle

%Table of Contents
\tableofcontents

\section{Introduction} \label{section: introduction}

Much of our understanding about chemistry at the microscopic scale is obtained through the studies of elementary reactions in the gas phase \cite{yang2007state,levine2009molecular}.  Since the pioneering crossed molecular beam experiments of Hershbach and Lee \cite{herschbach1987molecular}, refinements in experimental control over how such reactions are initiated have led to increasingly detailed understanding of their dynamics and mechanisms. Such advances are thanks in large part to developments in techniques to cool molecular samples substantially below room temperature, such that both the translational and integral degrees of freedom of the reactants become better-defined. For example, supersonic expansion resulted in significant narrowing in the velocity distribution of molecular beams, enabling the observation of dynamical features that depend sensitively on the collision energy such as quantized bottleneck transition states \cite{dai2003interference} and quantum interference between reaction pathways \cite{xie2020quantum}; beam merging and stark deceleration \cite{jankunas2015cold} have pushed the lower limit on achievable translational energy in molecular beams, allowing studies of reactions involving only a handful of partial waves where resolved collisional resonances associated with individual partial waves were observed \cite{lavert2014observation,de2020imaging}; vibrationally and rotationally cold molecules, combined with techniques to efficiently manipulate these degrees of freedom \cite{vitanov2017stimulated,mukherjee2011stark}, enabled the investigations of state-dependent reaction dynamics, revealing dynamical resonances \cite{wang2013dynamical} and stereodynamic preferences \cite{perreault2017quantum}. Furthermore, cold molecules served as a natural platform for investigating reactions at astrochemically relevant temperatures ranging from a few to a few tens of Kelvin \cite{smith2011laboratory}.

\begin{marginnote}[]
\entry{The ultracold regime}{A temperature regime which typically corresponds to millikelvins or below, where collisions between reactants are restircted to occur via the single lowest allowed partial wave of the system, and all internal quantum degrees of the freedom of the reactants are fully controlled.}
\end{marginnote}

\begin{figure}[t!]
\centering
\includegraphics[width=1.0\textwidth]{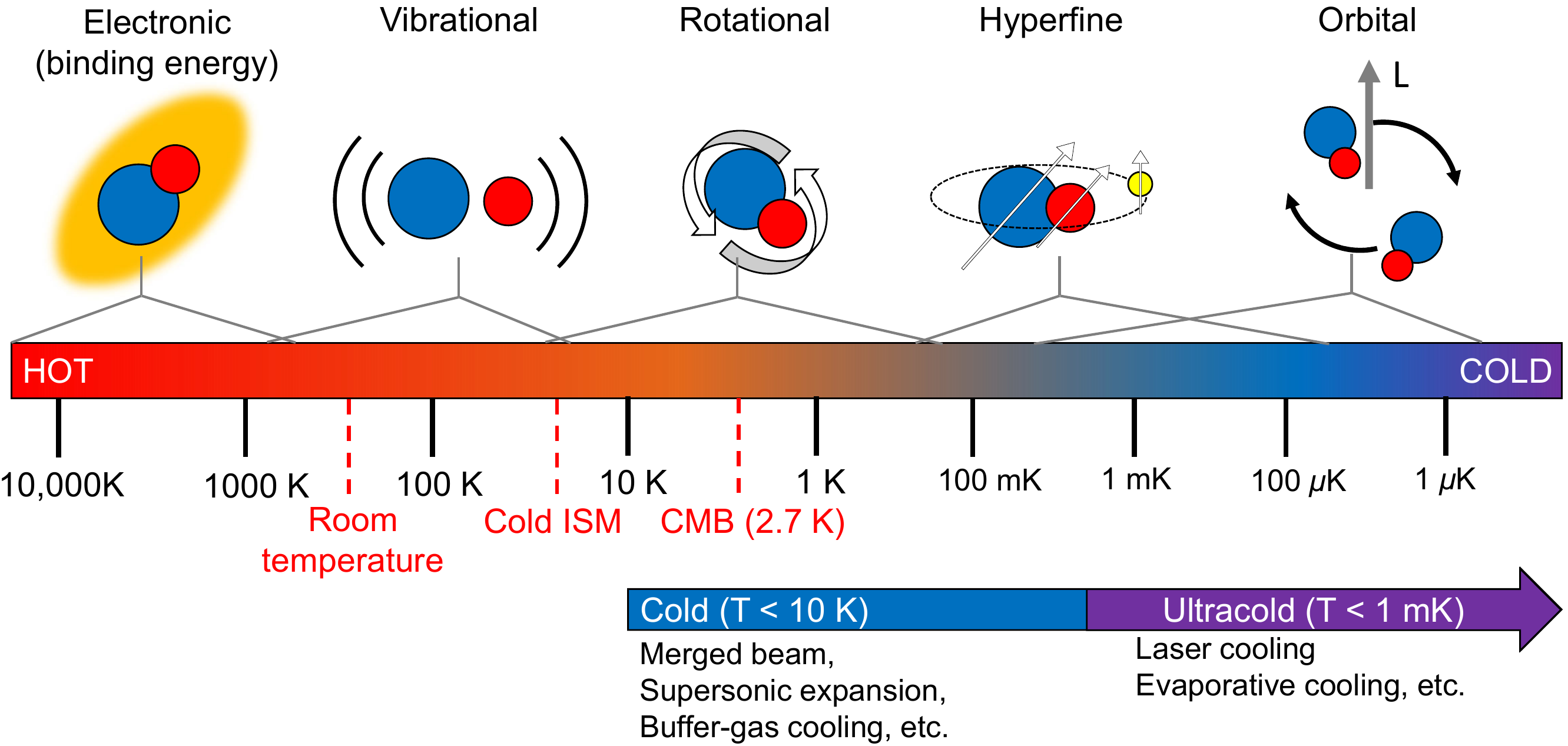}
    \caption{The molecular thermometer. Characteristic energy scales ($E$) of various degrees of freedom of a diatomic molecule are expressed in equivalent temperature values ($T$), according to $E = k_B T$, where $k_B$ is the Boltzmann constant. The energy scale for a given degree of freedom can span several orders of magnitudes due to the wide range of masses, intramolecular bindings, and intermolecular interactions associated with different molecules. While cold reaction temperature ($T \lesssim 10$ K) can be achieved in molecular beams, reaching the ultracold regime ($T \lesssim 1$ mK) requires AMO techniques such as laser cooling and evaporative cooling (section \ref{section: production}). ISM: interstellar medium. CMB: cosmic microwave background. }
\label{figMolecularThermometer}
\end{figure}

Given the significant advantages cooling has brought to the study of reaction dynamics, a natural question arises: how low in temperature does one need to go in order to control all quantum degrees of freedom within a molecule? The answer is illustrated in Fig. \ref{figMolecularThermometer}, where the typical energy scales for the various molecular degrees of freedom of a diatomic molecule are displayed on a ``molecular thermometer'' that spans ten orders of magnitude. Most molecules already exist in their electronic ground states at room temperature, and the vibrational and rotational ground states can be reached through established cooling techniques. However, further cooling is required to enter a single quantum state of the hyperfine degree of freedom, which describes the interactions between the spins of the electrons and the nuclei. Lower yet temperatures must be reached to fully control the orbital degree of freedom, such that collisions between reactants occur predominantly via the lowest allowed partial wave ($s-$wave for distinguishable particles or identical bosons; $p-$wave for identical fermions). This single partial wave regime is typically referred to as the ``ultracold'' regime, and occurs when the collision energy is much lower than the height of the centrifugal barrier associated with the next higher partial wave. Due to the wide range of particle masses and interaction strengths encountered in different reactions, the temperature threshold for the ultracold regime range from millikelvins to microkelvins \cite{zuchowski2010reactions}.

\begin{marginnote}[]
\entry{Energy unit conversion}{\\1 K is equivalent to \\
86.17 $\mu$eV \\
0.6950 cm$^{-1}$ \\
20.84 GHz \\
8.320$\times$10$^{-3}$ kJ/mol \\}
\end{marginnote}

\begin{figure}[t!]
\centering
\includegraphics[width=1.2\textwidth]{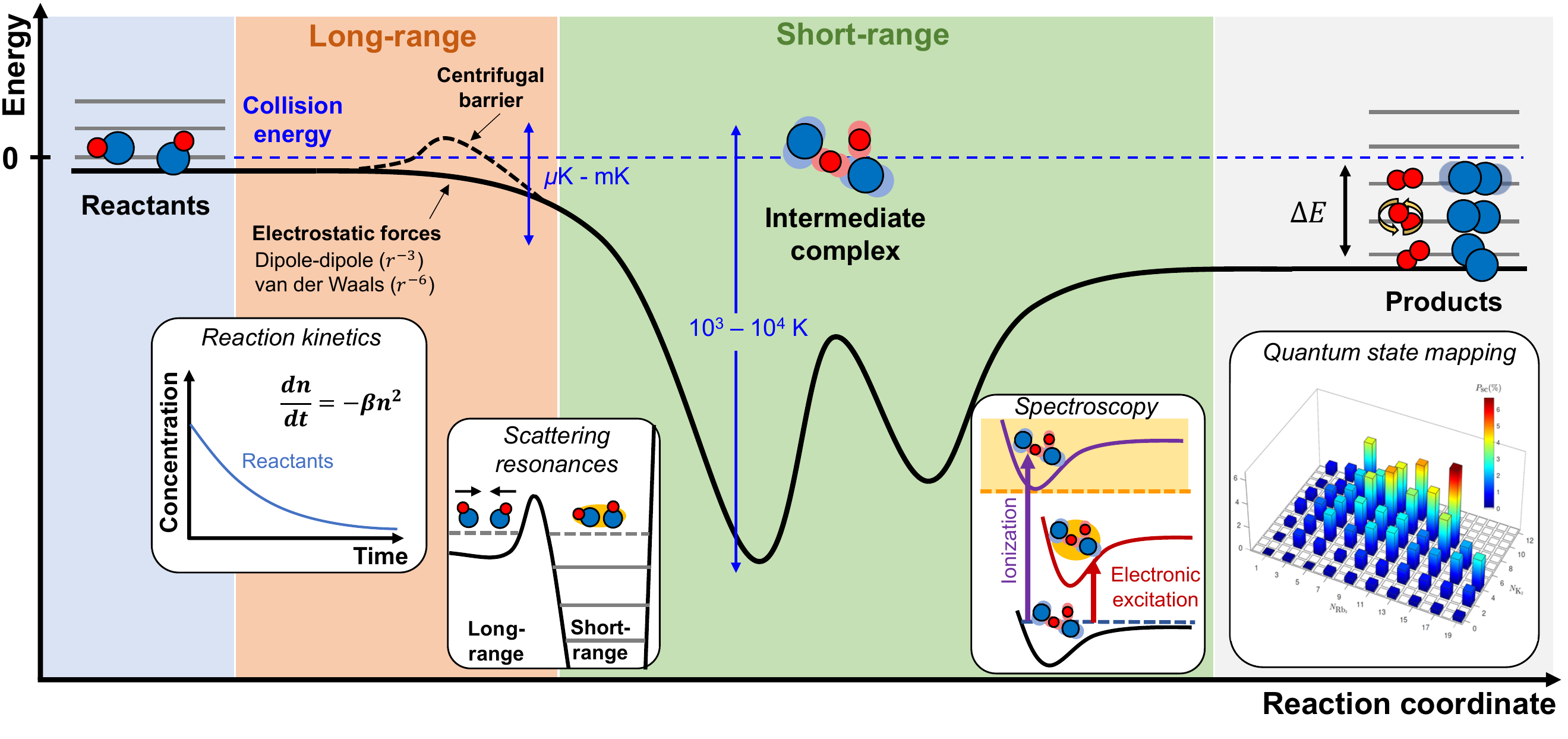}
    \caption{Schematic potential energy surface of a barrierless, exothermic reaction along some reaction coordinate, divided into long-range and short-range portions (not to scale). Experimental techniques used to probe the different portions are illustrated in white boxes. Figure adapted from Refs. \cite{liu2020probing} and \cite{liu2021precision} with permission.}
\label{figSchematicPES}
\end{figure}

Over the past decade, rapid progress in atomic, molecular, and optical (AMO) physics techniques have turned the dream of ultracold molecules into a reality, thereby opening the possibility to explore chemistry in a brand new temperature regime and with unprecedented control. We begin this article with a brief overview of experimental methods for creating and manipulating ultracold molecules. Then, we review recent progress in studying bimolecular chemistry at ultralow temperatures. Our discussion will be divided into two sections, respectively focusing on the long-range and short-range portions of the potential energy surface (PES) on which a reaction occurs (Fig. \ref{figSchematicPES}). Long-range forces, including van der Waals, dipole-dipole, and centrifugal, guide the approach between a pair of reactants. Once their distance approaches the order of a bond length, they reach the short-range where Coulomb interactions govern the rearrangements of the electrons and nuclei into products. Through example works, we highlight ways in which ultracold reactions behave differently from their higher temperature counterparts, and illustrate the remarkable degree to which they can be controlled. The scope of this review is limited to reactions involving neutral species, as the chemistry of cold and ultracold molecular ions has been extensively reviewed elsewhere \cite{zhang2017cold,toscano2020cold,heazlewood2021towards}.

\begin{marginnote}[]
\entry{AMO}{Atomic, molecular, and optical}
\entry{PES}{Potential energy surface}
\end{marginnote}

\section{Production, manipulation, and detection of ultracold molecules} \label{section: production}

%\sout{A necessary precursor to the realization of ultracold molecular samples is the development of atomic cooling techniques. In the 1980s, the application of laser cooling \cite{chu1989laser} allowed the temperature of certain neutral atomic species to be brought to below a milikelvin. Later on, magnetic trapping \cite{hess1987magnetic} and evaporation \cite{hess1986evaporative} enabled further cooling of the atoms into the microkelvin regime as well as increases in their density, which led to the creations of the first Bose-Einstein condensates \cite{anderson1995observation,davis1995bose}. These days, ultracold atoms are routinely created in experiments around the world, with wide-ranging applications from precision measurements to quantum computation.}

A necessary precursor to ultracold molecules is the maturation of techniques for creating and examining ultracold atoms, which have powered a wide range of applications in quantum science and engineering from precision measurements to quantum computation.
Compare to atoms, molecules display much richer internal structures that can be exploited to elevate many of these applications. For example, the strong and tunable electric dipole-dipole interactions \cite{ni2009dipolar} between molecules make them suitable building blocks for quantum simulation \cite{micheli2006toolbox,buchler2007strongly, cooper2009stable} and quantum computation \cite{demille2002quantum,yelin2006schemes,ni2018dipolar,hudson2018dipolar}; rovibrational transitions in certain classes of molecules offer high sensitivity to fundamental quantities such as the electron electric dipole moment \cite{baron2014order,cairncross2017precision,andreev2018improved} and the electron-to-proton mass ratio \cite{kokish2018prospects,kondov2019molecular}. As an added bonus, ultracold molecules offer a platform to explore chemistry with an unprecedented level of control \cite{balakrishnan2016perspective}.

At the same time, the internal structures of molecules pose a major challenge to their cooling. Laser cooling has been the indispensable first step to ultracold atom experiments \cite{chu1989laser}. This technique relies on rapid and repeated scattering of photons between two levels within the atom (typically the ground and first electronically excited states) to slow its motion to a near stop. While good photon-cycling properties can be found in many atomic species, most notably alkali and alkaline earth atoms, they are in general not available in molecules. Molecules, once electronically excited, can decay via spontaneous emission into a wealth of vibrational states on the ground electronic surface, which renders it unavailable for re-excitation by the same laser.

\begin{marginnote}[]
\entry{STIRAP}{stimulated Raman adiabatic passage}
\end{marginnote}

\begin{figure}[t!]
\centering
\includegraphics[width=1.2\textwidth]{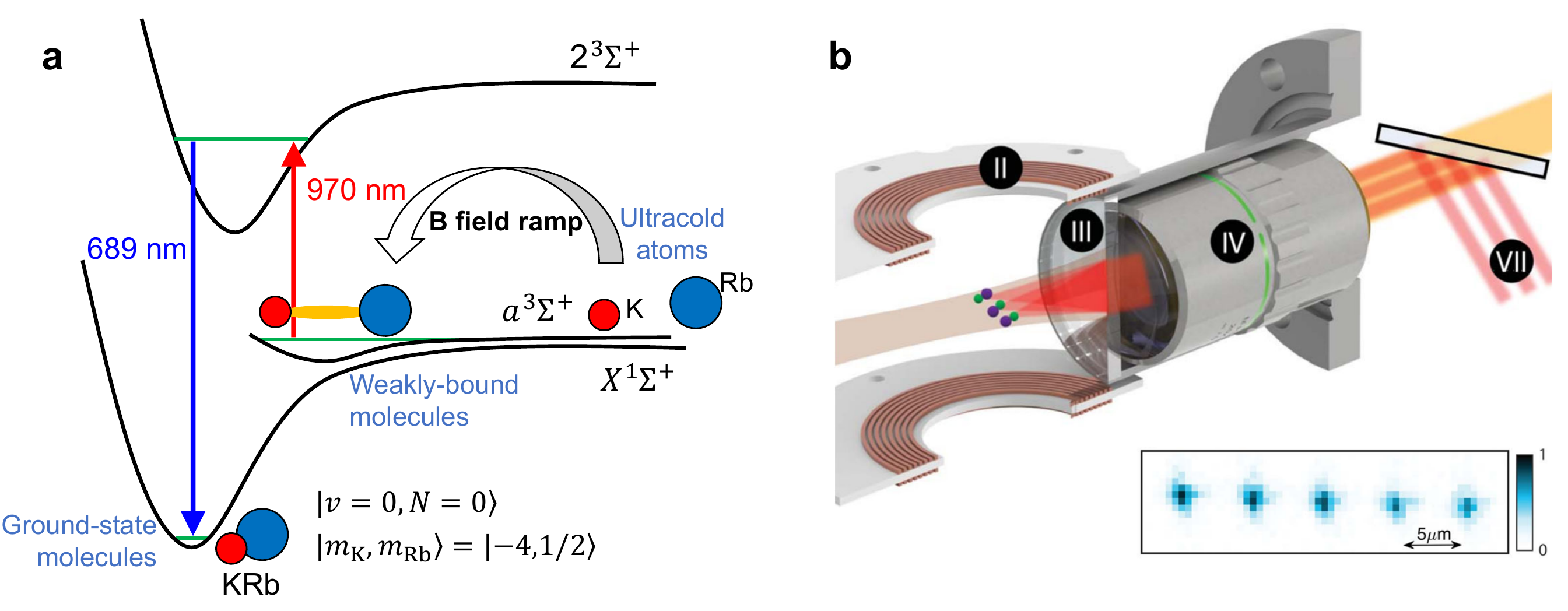}
    \caption{Creation and trapping of ultracold molecules. (a) Association of ultracold molecules from ultracold atoms. In this scheme demonstrated by Ni \textit{et al.} \cite{ni2008high}, ultracold K and Rb atoms are first converted into weakly-bound molecules by ramping an external magnetic field across an interspecies Feshbach resonance location at 546.7 G; then, the molecular population is transferred to a select hyperfine state within the rovibronic ground state using a STIRAP transition. (b) An optical tweezer array of ultracold CaF molecules. Each individual tweezer is formed by focusing a collimated laser beam (red) down to near the diffraction limit using a high numerical aperture objective. Fluorescence from the molecules is collected through the objective and imaged onto a camera. Inset: image of optical tweezer array of single molecules, averaged over 500 experiments. Part (b) of this figure is adapted from Ref. \cite{anderegg2019optical} with permission.}
\label{figCoolingAndTrapping}
\end{figure}

To circumvent this issue, two general approaches emerged -- one is to find molecules that can be directly laser-cooled, while the other is to associate laser-cooled atoms to form molecules. The direct cooling scheme exploits the exceptionally high population branching back to the original state predicted for certain classes of molecules \cite{gao2014laser}, including alkaline earth halides and group 3 metal oxides. Thus far, it has been demonstrated on SrF \cite{shuman2010laser}, CaF \cite{zhelyazkova2014laser,anderegg2018laser}, YO \cite{ding2020sub}, and has recently been extended to polyatomic species including SrOH \cite{kozyryev2017sisyphus}, CaOH \cite{baum20201d} and CaOCH$_3$ \cite{mitra2020direct}. In the atom association scheme, pairs of ultracold atoms are first magneto-associated into weakly-bond Feshbach molecules, and then transferred into the rovibronic ground state via stimulated Raman adiabatic passage (STIRAP), a coherent two-photon transition (Fig. \ref{figCoolingAndTrapping}a). The Feshbach molecule acts as an intermediary between the free atom pair and the ground state molecule, bridging the enormous differences between their bond lengths and binding energies. This technique was first demonstrated on homonuclear molecules by Lang \textit{et al.}\cite{lang2008ultracold} (Rb$_2$) and Danzl \textit{et al.}\cite{lang2008ultracold} (Cs$_2$), and on heteronucler molecules the JILA KRb team \cite{ni2008high} (KRb). Recent years saw an explosion in the number of new ultracold bilakali species, including RbCs~\cite{takekoshi2014ultracold,molony2014creation}, NaK~\cite{park2015ultracold,rui2017controlled,Voges2020,bause2021collisions}, NaRb~\cite{guo2016creation}, LiNa~\cite{rvachov2017long}, LiK~\cite{Yang2020}, and NaCs~\cite{cairncross2021assembly}. For further information on these techniques, we refer the readers to Refs. \cite{tarbutt2019laser} and \cite{ni2009dipolar}. Once close to or within the ultracold regime, the temperature of a molecular sample can be further reduced through evaporative cooling \cite{son2020collisional,valtolina2020dipolar}.

\begin{marginnote}[]
\entry{ODT}{optial dipole trap}
\entry{Optical tweezer}{An ODT with a diffraction-limited focus designed to hold a single particle}
\end{marginnote}

Once the molecules are ultracold, they can be trapped using methods developed for ultracold atoms, enabling extended observation and further manipulation. Owing to its simplicity and versatility, optical dipole trap (ODT) has been the preferred ``beaker'' of choice among ultracold molecule experiments \cite{grimm2000optical}. In the most basic version of this scheme, a Gaussian laser beam whose frequency is far red-detuned from one or more transitions within the molecule is applied to the sample, exerting an optical dipole force that attracts the molecules towards its focus. Using multiple beams in specific spatial arrangements creates different trap geometries, including two-dimensional layers \cite{ni2010dipolar} and three-dimensional lattice structures \cite{danzl2010ultracold,moses2015creation}. Furthermore, an ODT with a near-diffraction-limited waist can be used to hold precisely one molecule. Such a configuration is known as an optical tweezer \cite{dumke2002micro}, and has been recently demonstrated on NaCs \cite{liu2018building, zhang2020forming,yu2020coherent, cairncross2021assembly}, Rb$_2$ \cite{he2020coherently}, and CaF \cite{anderegg2019optical} molecules (Fig. \ref{figCoolingAndTrapping}b). This technique is of significant interest for ultracold bimolecular chemistry: by merging two tweezers, each holding a reactant molecule, reactions can be probed one event at a time, free of any ensemble averaging \cite{cheuk2020observation}. Recently, however, it was discovered that the intense trapping light can strongly influence the dynamics of certain ultracold reactions (section \ref{subsection: long-lived complex}). To this end, static field traps, which confine molecules through their intrinsic electric or magnetic dipole moments, may be viable alternatives \cite{sawyer2007magnetoelectrostatic,mccarron2018magnetic}.

While ultracold molecules typically exist in the rovibronic ground state at the end of their cooling, they can be transferred to the desired state using well-established AMO techniques. For example, microwave pulses are routinely used to drive molecules between ground and first-excited rotational states, or to change their hyperfine states \cite{will2016coherent}; Raman pulses have been employed to prepare molecules in vibrationally excited states \cite{ye2018collisions,kondov2019molecular}. Static magnetic (electric) fields can be applied to mix different spin (rotational) levels within a molecule, aligning its magnetic (electric) dipole moment in the lab frame \cite{hu2020nuclear,ni2009dipolar}. Recently, optical frequency combs have emerged as a versatile tool for molecular state control \cite{chou2020frequency}. Boasting both high spectral bandwidth and coherence, they may be used to coherently drive different transitions in a wide range of molecular species. 
%\sout{Thanks to the exceptional spectral resolutions achievable at ultralow temperatures, the efficiencies of state manipulations in ultracold molecules can oftentimes approach unity.}

Detection of the ultracold molecular population at the end of an experiment is typically achieved through optical imaging \cite{ketterle1999making}, with the two most common examples being absorption and fluorescence imaging (Fig. \ref{figMachine}a\&b). Optical images provide direct visualizations of the spatial configurations of particles and information such as their number, density, momentum, and temperature. Much like in laser cooling, rapid photon-cycling is required to detect ultracold molecules, which typically exist in small quantities ($<10^5$ particles) and low density ($10^{8} - 10^{14}$~\textrm{particles}/cm$^{3}$). To this end, molecules produced by direct laser cooling can be imaged by scattering photons on the same optical transition used for cooling \cite{tarbutt2019laser}; those associated from ultracold atoms can be converted back into atoms through a reversal of the association process, which can then scattering photons efficiently \cite{ni2009dipolar}. The sensitivity of optical imaging to single quantum states of molecules is a vital resource for studying how ultracold reactivity depend on different states of the reactants, which is the topic of section \ref{section: long-range}. At the same time, this high degree of specificity limits the usefulness of imaging for detecting other chemical species involved in reactions, including the intermediate complex and the reaction products, which exist in a large number of quantum states. As we will show in section \ref{section: short-range}, overcoming this limitation through physical chemistry techniques has led to significant improvements in our ability to probe and understand ultracold reactions.

\section{Ultracold molecular chemistry: the long-range} \label{section: long-range}

Even before the start of significant experimental efforts in ultracold chemistry, it was recognized that long-range interactions, \textit{e.g.} van dar Waals, dipole-dipole, and centrifugal, can profoundly influence reactivity in the ultracold regime \cite{weck2006importance,bell2009ultracold} (Fig. \ref{figSchematicPES}). Here, the energy scale of these relatively weak interactions become comparable to or even greater than the collision energy. As such, different forms of long-range potential can result in dramatically different reaction rates. Furthermore, since both the energy and the impact parameters become severely restricted for ultracold collisions, collisional resonances become resolvable. The position of such resonances depend sensitively on both the long and short-range potentials, and their presence can drastically enhance the reaction rate over non-resonant cases. These unique properties led to a flurry of activities in studying the role of long-range interactions in ultracold reactions over the past decade. Our review in this section shall focus primarily on recent studies, as excellent reviews of earlier works can be found elsewhere \cite{dulieu2009formation, nesbitt2012toward,quemener2012ultracold,balakrishnan2016perspective,bohn2017cold}.

Ultracold chemical reactivity was first observed by the JILA KRb team in a sample of ultracold KRb molecules \cite{ospelkaus2010quantum}, which are among the first species to be brought into the new temperature regime. In this landmark study, Ospelkaus \textit{et al.} observed rapid and unexpected losses in a gas of $^{40}$K$^{87}$Rb molecules prepared in the rovibronic ground state. These losses were attributed to the bimolecular reaction KRb + KRb $\rightarrow$ K$_2$ + Rb$_2$. Remarkably, the reaction rate displayed a strong dependence on whether the KRb molecules are prepared in identical or statistically-mixed hyperfine states, with a 10 -- 100 fold difference observed between the two cases. The explanation for this effect lies in symmetries imposed by quantum statistics: identically-prepared molecules are restricted to react via $p-$wave collisions due to the fermionic nature of $^{40}$K$^{87}$Rb, resulting in the presence of a centrifugal barrier that limits the transmission probability of the reactants into the short-range; those in a statistical mixture, on the other hand, are distinguishable and can therefore react via barrierless $s-$wave collisions, resulting in much faster losses.

The KRb study demonstrated that the long-range potential of an ultracold reaction is directly tied to the internal states of the reactants, which can be efficiently manipulated using AMO techniques. 
Dipolar molecules represent an ideal platform to further explore this controllability, as the dipole-dipole interactions between the reactants depend on the quantum states of their rotations, which can be controlled using either static or time-varying (\textit{i.e.} microwave) external electric fields.

In the presence of a static field, ultracold molecules align in the lab-frame due to a mixing of opposite-parity rotational states, and their interactions become anisotropic. This effect was explored by the JILA KRb team under different confinement geometries for the ultracold sample. For a sample trapped in an ODT, where collisions can occur in all three spatial dimensions, it was found that the attractive interactions between reactants aligned head-to-tail lowers the previously observed $p$-wave barrier, leading to a significant enhancement of the overall reaction rate \cite{ni2010dipolar}. For a sample trapped in a one-dimensional optical lattice, on the other hand, collisions are restricted to occur within pancake-like layers in a repulsive side-by-side configuration, leading to a suppression of the chemical reaction \cite{de2011controlling}. Guo \textit{et al.} studied dipolar collisions between ultracold bosonic NaRb molecules, and found that large dipolar interactions lead to couplings between different participating partial waves, resulting in a stepwise enhancement in the two-body loss rate as a function of the lab-frame dipole moment of NaRb.

\begin{figure}[t!]
\centering
\includegraphics[width=1.0\textwidth]{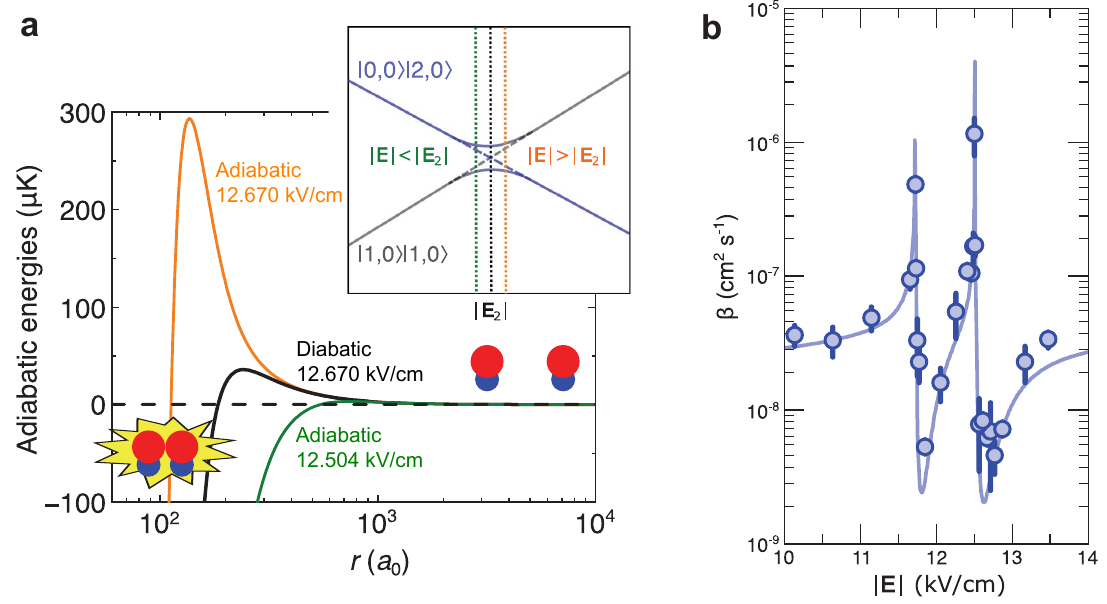}
    \caption{Shielding an ultracold reaction using static electric field. (a) Adiabatic potential energy curves for KRb molecules in the rotational state $| J = 1, m_J = 0\rangle$ colliding via a side-by-side geometry under an electric field with strength $|\mathbf{E}|$ that slightly above (solid orange) or below (solid green) $|\mathbf{E_2}| = 12.51 $ kV/cm. $r$ represents the separation between the colliding molecules. (inset) At $|\mathbf{E_2}|$, resonant coupling between the states $|1,0\rangle|1,0\rangle$ and $|0,0\rangle|0,2\rangle$ creates an avoided crossing, resulting in an effective attraction when the molecule collide at $|\mathbf{E}| < |\mathbf{E_2}| $ (dotted orange line) and repulsion at $|\mathbf{E}| > |\mathbf{E_2}| $ (dotted orange line). The solid black curve represents the diabatic potential curve, where the effect of the resonant coupling is absent. The dashed black line indicates the average collisional energy of 250 nK. (b) The two-body loss rate of KRb molecules ($\beta$) measured at various electric field strengths. In a window of $\sim 0.5$ kV/cm round $|\mathbf{E_2}|$, nearly three orders of magnitude of variation in $\beta$ is observed. Another resonant feature is observed at $|\mathbf{E}| = 11.72$ kV/cm, which arises from the coupling between $|1,0\rangle|1,0\rangle$ and $|0,0\rangle|2,\pm1\rangle$. Figure adapted from Ref. \cite{matsuda2020resonant} with permission.}
\label{figEFieldShielding}
\end{figure}

As an alternative to static electric fields, a resonant microwave drives can be used to directly couple opposite parity rotational states within ultracold molecules, resulting in strong dipolar interactions. Yan \textit{et al.} \cite{yan2020resonant} studied the collisions between microwave-dressed ultracold $^{23}$Na$^{40}$K molecules, where the field near-resonantly couples the ground ($|J = 0, m_J = 0 \rangle$) and first excited ($| 1, 1 \rangle$) rotational states, and found the two-body collision cross section to be several orders of magnitude higher than the background cross section. This was attributed to a substantial lowering of the $p$-wave barrier experienced by the fermionic molecules due to the strong dipole-dipole interactions induced by the dressing. He \textit{et al.} \cite{he2021observation} studied resonant dipolar collisions between ultracold NaRb molecules in different rotational states. By applying a microwave $\pi/2$ pulse to a sample of ground state molecules and waiting for a short period for the coherence in population to decay, they prepared a statistical mixture of $J = 0$ and $J = 1$ molecules. The ground-excited pair can then exchange excitation resonantly as they collide, resulting in a strong dipole-dipole interactions that increase the collision cross section over non-dipolar cases by up to 20 times.

While attractive dipolar interactions can enhance the rates of ultracold reactions, repulsive interactions can be used to suppress such reactions when they are undesired. Using a static electric field, Matsuda \textit{et al.} \cite{matsuda2020resonant} achieved highly effective shielding of reactions between ultracold KRb molecules prepared in the rotational state $| J = 1, m_J = 0\rangle$ and confined in quasi two-dimensional geometry. It was found that at a specific field value of 12.51 kV/cm the energies of a pair of molecules in the state $| 1 , 0 \rangle | 1, 0 \rangle$ becomes degenerate with those in $| 0 , 0 \rangle | 2, 0 \rangle$, and the resonant dipolar interaction between these states gives rise to an avoided crossing. At field values slightly above 12.51 kV/cm, the avoided crossing leads to a strong repulsive interactions between the molecules that suppresses the two-body loss rate by an order of magnitude over the non-resonant cases (Fig. \ref{figEFieldShielding}). Subsequently, this shielding mechanism was extended to all three spatial dimensions \cite{li2021controlling}. Another approach of shielding collisions in three dimensions was recently demonstrated by Anderegg et al.~\cite{anderegg2021observation} for a pair of ground-state CaF molecules confined in an optical tweezer using microwave dressing.  It was shown that a microwave field blue-detuned from the transition between the rotational ground and first-excited states creates an effective repulsive interaction between the colliding pair, reducing the rate of two-body losses by a factor of six.

In recent years, scattering resonances has emerged as an important subject of research in ultracold molecular chemistry. A resonance arises when a scattering state of the reactants or products becomes degenerate in energy with and strongly coupled to a bound or quasi-bound state of the intermediate complex (see inset to Fig. \ref{figSchematicPES}). Resonances involving quasi-bound states are known as ``shape resonances'', while those involving true bound states are known as ``Feshbach resonances''. 
%\sout{While the observation of scattering resonances under thermal conditions is typically hindered by effects of ensemble averaging, it becomes possible in low temperatures reactions where collision energies are low and narrowly distributed.}
Shape resonances have been extensively studied in the cold regime using crossed or merged molecular beams \cite{jankunas2015cold}. Recently, de Jongh \textit{et al.} pushed the accessible collision energy of crossed molecular beams into the ultracold regime using Stark deceleration. Using the state-of-the-art apparatus, they observed shape resonances in collisions between He and NO that are dominated by $s-$ and $p-$waves. The measured resonance energies that could only be reproduced at the CCSDT(Q) level, challenging the accuracy of the CCSD(T) method widely consider to be the ``gold standard'' in electronic structure calculations. Feshbach resonances play important roles in ultracold collisions of atoms~\cite{chin2010feshbach}, and 
%\sout{have been predicted for both atom-molecule \cite{mayle2012statistical,hummon2011cold,frye2016approach} and molecule-molecule \cite{bohn2002rotational,tscherbul2009magnetic} collisions. Recently, magnetically tunable Feshbach resonances were}
 were recently observed in ultracold collisions between NaK and K \cite{yang2019observation}, setting a valuable benchmark for quantum chemistry calculations of heavy atom systems.

\section{Ultracold molecular chemistry: the short-range} \label{section: short-range}

Once the colliding partners traverse through the long-range potential, they reach the short-range where the exchange of particles occur (Fig. \ref{figSchematicPES}). Exploring the detailed dynamics of this exchange, both experimentally \cite{yang2007state} and theoretically \cite{clary2008theoretical}, has been central to our understanding and description of chemistry at the quantum level, and continues to be at the forefront of physical chemistry research \cite{zhang2016recent,li2020advances}. The ability to initiate a reaction in the ultracold regime brings several unique opportunities to the studies of short-range reaction dynamics. Since all quantum degrees of freedom of the reacting molecules are completely defined, one could study reactions at a truly state-to-state level. The resulting product state distribution will provide the ultimate benchmark for dynamical theories, and may reveal quantum effects (\textit{e.g.} resonances, interference, and entanglement) that are typically obscured by thermal averaging over many reactant states. Furthermore, the quantum states of the ultracold reactants may be manipulated to influence the pathway and outcome of the reaction, enabling state-to-state control of chemistry with greatly improved precision. Finally, exploring the ways in which the short-range dynamics of ultracold reactions are similar to or different from their``hotter'' counterparts will expand our description of chemistry to a wider range of starting temperatures.

\begin{figure}[t!]
\centering
\includegraphics[width=1.2\textwidth]{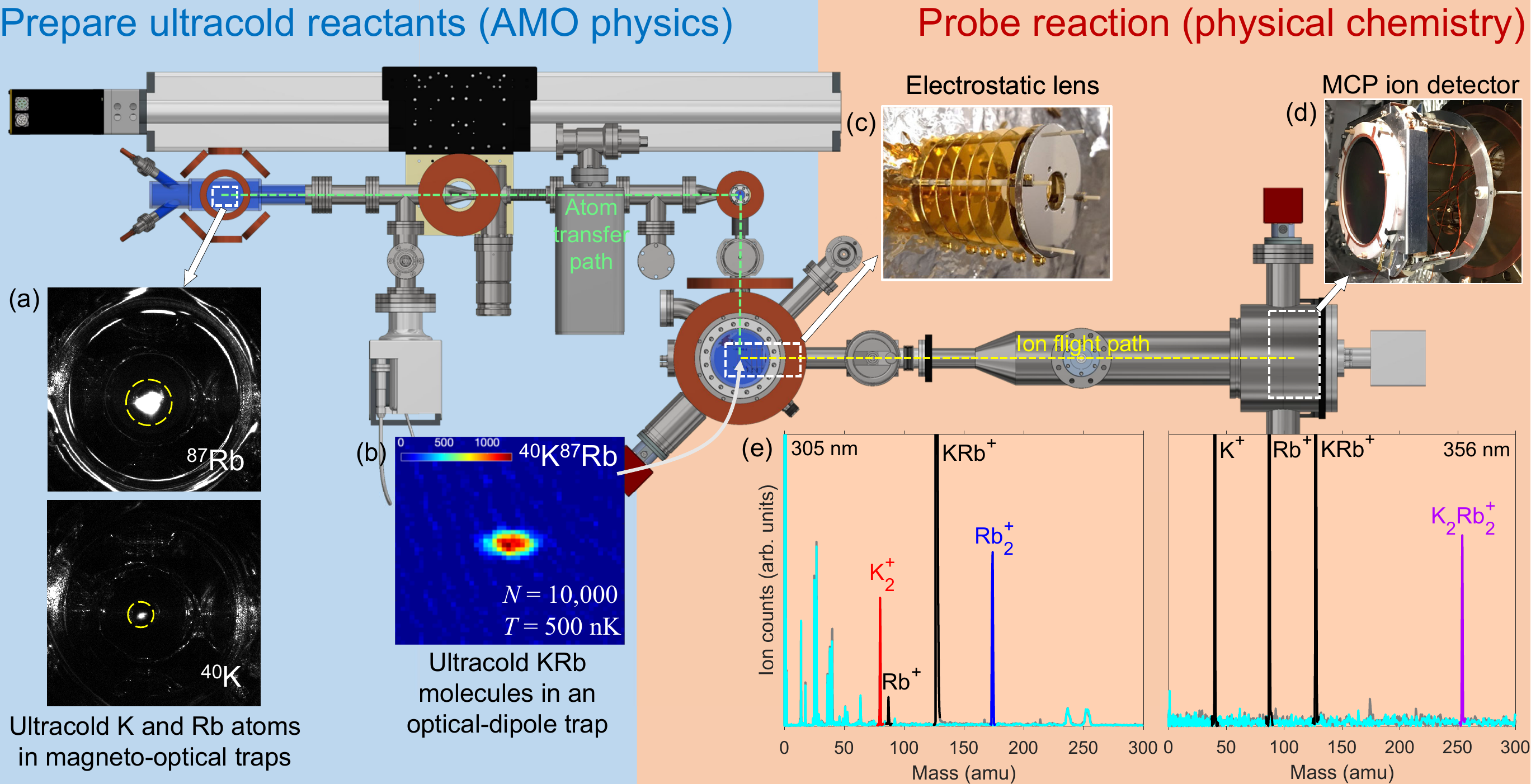}
    \caption{The Ultracold chemistry machine at Harvard. The blue shaded portion utilizes AMO techniques (\textit{i.e.} laser cooling, coherent state-preparation) to create state-controlled reactants at ultralow temperatures, while the red shaded portion employs physical chemistry techniques (\textit{i.e.} ion spectrometry) to probe the ensuing reactions. Insets: (a) Fluorescence images of laser-cooled $^{87}$Rb and $^{40}$K atoms confined in magneto-optical traps. (b) Absorption image of a cloud of KRb molecules at a temperature of 500 nK. (c) Ion optics for velocity-map imaging. (d) Multi-channel plate (MCP) ion detector. (e) Ion TOF spectra obtained by performing photoionization on a sample of KRb moleucles undergoing two-body reactions; peaks associated with the reactants (KRb), the products (K$_2$,Rb$_2$), and the intermediate complex (K$_2$Rb$_2^*$) are observed. Figure adapted from Ref. \cite{liu2020probing} with permission.}
\label{figMachine}
\end{figure}

Despite the success in understanding the long-range characteristics of ultracold molecular reactions over the past decade (section \ref{section: long-range}), experimental explorations of the short-range have remained scarce. In many ultracold chemistry studies, the rate at which reactants disappear, which can be conveniently measured through optical imaging (section \ref{section: production}), serves as the sole experimental observable. While such kinetics measurements can reveal the role the long-range potential plays in deciding how the reactants come together, they generally provide limited information on the subsequent short-range chemistry. To gain insights into the latter, one must directly probe the reaction dynamics or its aftermath. This requires the detection of the transient intermediates or the products of the reaction, for which optical imaging is ill-suited due to its high degree of specificity. On the other hand, a wealth of universal yet sensitive physical chemistry techniques have been developed for this very task thanks to decades of research in gas phase reaction dynamics. Laser-induced fluorescence\cite{greene1983determination} and state-selective ionization \cite{zhou2003mode} provided sensitive probes for the internal states of the products, while ion imaging \cite{eppink1997velocity,ashfold2006imaging} enabled measurements of their translational energies and angular distributions. Even the transient species involved in the reaction can be directly accessed. For example, timescales for the dissociation of intermediate complexes can be probed using ultrafast lasers \cite{zewail2000femtochemistry}, while structures of the transition states or complexes can be obtained through photodetachment \cite{garand2008nonadiabatic, continetti2017dynamics}, photoabsorption \cite{su2013infrared}, photodissociation \cite{sato2001photodissociation}, or Coulomb explosion \cite{vager1989coulomb}.

Thus, a comprehensive investigation of ultracold chemistry -- including their short-range dynamics -- calls for a combination of AMO techniques for producing ultracold molecules, and physical chemistry techniques for probing the process and outcome of their reactions. Such an integration was recently realized by the authors at Harvard \cite{liu2020probing}, as depicted schematically in Fig. \ref{figMachine}. Using this apparatus, we tracked an ultracold reaction from start to finish for the first time \cite{hu2019direct}. Here, a sample of KRb molecules were identically prepared in their absolute quantum ground states at a temperature of 500 nK using the coherent association technique demonstrated by Ni \textit{et al.} in 2008 \cite{ni2008high}; their subsequent two-body exchange reactions were probed by first photoionizing any species involved in the reaction, and then performing mass spectrometry and velocity map imaging (VMI) on the resulting ions \cite{liu2020probing}. This not only yielded ion signals associated with the reaction products, K$_2$ and Rb$_2$, but also that associated with the intermediate complex, K$_2$Rb$_2^*$ (Fig. \ref{figMachine}(e)). These direct signals enabled further investigations of the short-range dynamics of the ultracold KRb + KRb reaction, and allowed us to gain control over its various aspects. In this section, we review several key findings from these studies as well as related works from other groups.

\begin{marginnote}[]
\entry{VMI}{Velocity map imaging}
\entry{RRKM}{Rice-Ramsperger-Kassel-Markus}
\entry{DOS}{Density of states}
\end{marginnote}

\subsection{Long-lived intermediate complex} \label{subsection: long-lived complex}

The KRb + KRb reaction is an example of complex forming reactions -- an important class of reactions that is widely found in interstellar, atmospheric, and combustion processes \cite{guo2012quantum,osborn2017reaction}. As such, intermediate complexes have been extensively studied since the early days of chemistry \cite{light1967statistical,herschbach1973reactive,troe1994polanyi}, and continue to be an important subject of research today \cite{su2013infrared,bjork2016direct}.
In gas phase chemistry, intermediate complexes are noted by their fleeting nature -- typically lasting on the order of 10 ps or less in systems consisting of a few atoms. The investigations of these highly transient species have relied on two general types of methods: follow the complexes' motions using ultrafast lasers \cite{zewail2000femtochemistry}, or arrest their motions through collisional stablizations \cite{womack2015observation,bjork2016direct}.

\begin{figure}[t!]
\centering
\includegraphics[width=1.2\textwidth]{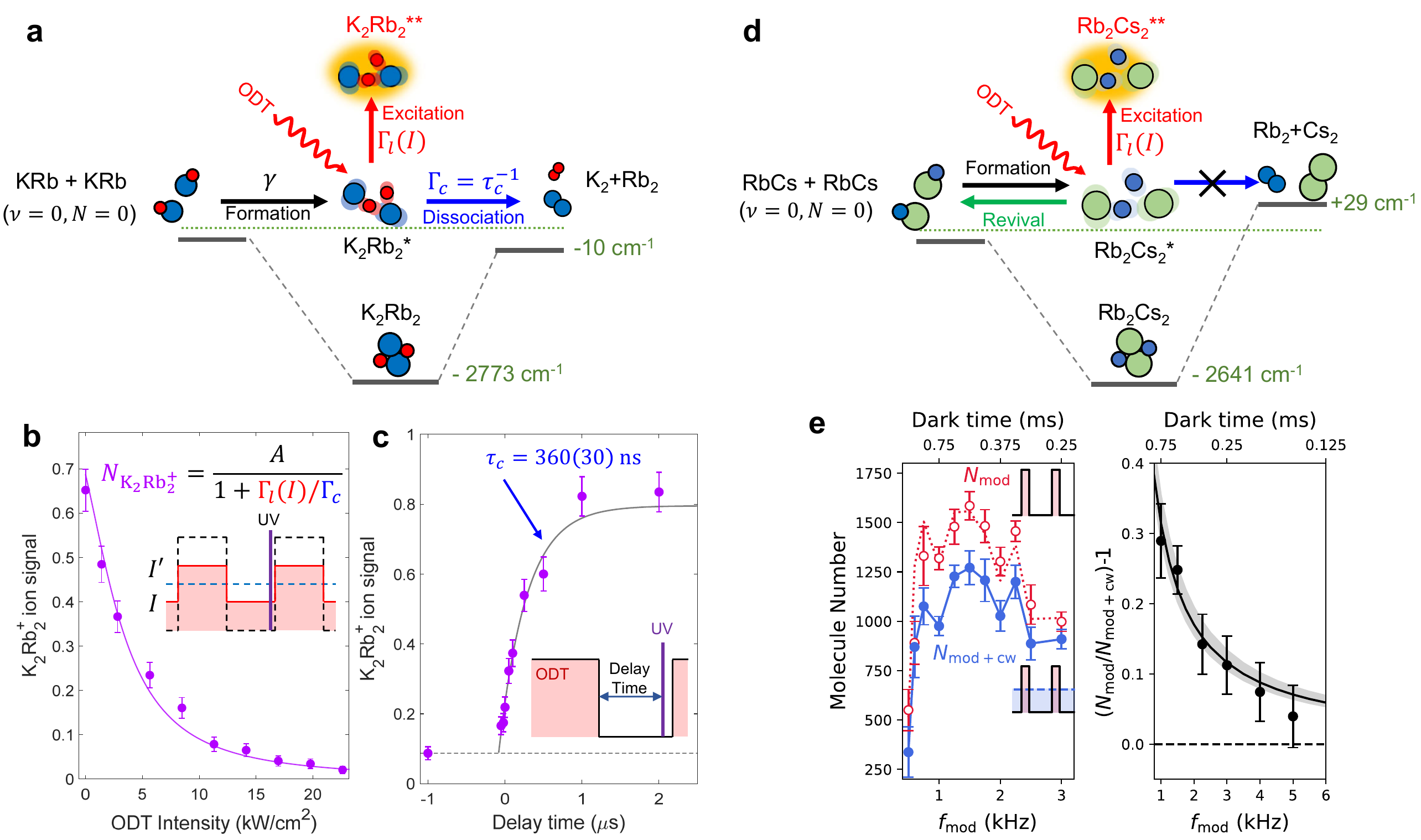}
    \caption{Long-lived intermediate complexes in ultracold reactions. (a) Energetics of the exothermic KRb + KRb reaction. The intermediate complex K$_2$Rb$_2^*$, formed from collisions between two KRb molecules, can either dissociate into products or undergo photoexcitation by the trap light. $\Gamma_c$: dissociation rate; $\tau_c$: complex lifetime; $\Gamma_l$: excitation rate; $I$: trapping light intensity. (b) K$_2$Rb$_2^+$ signal as a function of the ODT intensity during the $I$-phase of the intensity modulation (inset). (c) Measuring the complex lifetime by monitoring growth in the complex population following a sudden switch-off of the ODT (inset). (d) Energetics of the endothermic RbCs + RbCs reaction. The Rb$_2$Cs$_2^*$ complex can either revive back into two RbCs molecules or undergo photoexcitation. (e) Evidence for the photoexcitation of Rb$_2$Cs$_2^*$ by 1550 nm light. Red series: the number of RbCs molecules remaining in the sample after a fixed period of confinement in a time-modulated 1064 nm ODT ($N_{\rm{mod}}$) as a function of the ODT modulation frequency ($f_{\rm{mod}}$); blue series: the same quantity measured in the presence of an additional CW 1550 nm light ($N_{\rm{mod + CW}}$); black series: fractional difference between $N_{\rm{mod}}$ and $N_{\rm{mod + CW}}$. The suppression of the complex revival at increasing modulation frequency is due to the shortening of the dark periods of the modulation during which revival is allowed. Figure adapted from Refs. \cite{liu2020photo} and \cite{gregory2020loss} with permissions.}
\label{figLongLivedComplex}
\end{figure}

Notably, our detection of the transient K$_2$Rb$_2^*$ complex required neither of the above techniques \cite{hu2019direct}. Instead, by simply shinning a nanosecond-scale pulsed UV laser at the reacting cloud of KRb molecules, we observed rather efficient conversion of K$_2$Rb$_2^*$ into detectable K$_2$Rb$_2^+$ ions. This indicates that the lifetime of the K$_2$Rb$_2^*$ complex is substantially longer compared to those typically found in reactions of similar system sizes. We can qualitatively rationalize this long lifetime under the framework of the celebrated Rice-Ramsperger-Kassel-Markus (RRKM) statistical theory \cite{levine2009molecular}, which states that given complete energy mixing within the complex, its lifetime can be calculated as $\tau_{\textrm{c}} = h \rho(E_0)/N_0(E_0)$, where $\rho(E_0)$ represents the density of states (DOS) of the complex at the incident energy $E_0$, and $N_0(E_0)$ is the number of quantum states available for the complex to dissociate, \textit{i.e.} exit channels. By preparing the reactants in their absolute quantum ground states and at ultralow collisional temperatures, we achieve an exceptionally low $E_0$ that energetically minimizes $N_0$. Under such a condition, the deep potential well of the reaction ($\sim$ 3000 cm$^{-1}$) and the comparatively low exogerticity ($\sim$ 10 cm$^{-1}$) contribute to a large $\rho$ a small $N_0$, respectively, resulting in a quite substantial $\tau_{\textrm{c}}$ (Fig. \ref{figLongLivedComplex}a).

Long-lived complexes have been the subject of considerable theoretical interest within the AMO community due to their potential role in mediating population losses observed in ultracold molecular gases assumed to be chemically stable. Since the loss of KRb molecules was attributed to exothermic chemical reactions \cite{ospelkaus2010quantum}, many groups interested in harnessing the power of ultracold molecules for applications besides chemistry created bialkali species AB for which the reaction AB + AB $\rightarrow$ A$_2$ + B$_2$ is endothermic (Fig. \ref{figLongLivedComplex}d), including RbCs\cite{takekoshi2014ultracold}, NaK\cite{park2015ultracold}, and NaRb\cite{guo2016creation}. However, two-body losses at near-universal rates were still measured in all cases\cite{ye2018collisions,gregory2019sticky}, and it was suspected that transient complexes, A$_2$B$_2^*$, formed upon the pairwise collisions of AB, was the culprit. In a 2013 paper, Mayle \textit{et al.} put forth the "sticky collision" hypothesis, where the A$_2$B$_2^*$ complex, once formed, can live for long enough (\textit{e.g.} 45 ms in the case of Rb$_2$Cs$_2^*$) to collide ineslatically with another molecule within the sample, resulting in the ejection of both from the trap. 
%\sout{Using the ultracold RbCs + RbCs system as an example, they estimted a complex lifetime of 45 ms -- comparable to the average time it would take for the complex to encounter another molecule in the trap, which is $\sim$ 10 - 100 ms for molecule densities of 10$^{11}$ - 10$^{10}$ cm$^{-3}$.}
Recently however, Christanen \textit{et al.}\cite{christianen2019quasiclassical} estimated lifetimes for similar bialkali complexes to be $\sim$3 orders of magnitude shorter, making a secondary collision highly unlikely. Instead, it was proposed that the observed losses are due to electronic excitations of the complexes caused by the optical dipole trap in which the molecules are confined\cite{christianen2019photoinduced}. 
%\sout{Their calculations not only show the existence of excitation pathways for A$_2$B$_2^*$ at the commonly used ODT wavelengths of 1064 and 1550 nm, but also that the rate for such excitations can be orders of magnitude faster than the rate at which the complex dissociates back into AB + AB at the ODT intensity used during experiments, resulting in the apparent two-body loss.}

\begin{marginnote}[]
\entry{Universal two-body loss}{loss that occurs with unit probability once the colliding pair reach the short-range.}
\end{marginnote}

This unexplained loss remained a mystery for nearly a decade, and the direct ion signal on the intermediate complex puts us in a unique position to investigate it. In Ref. \cite{liu2020photo}, we studied the ultracold KRb + KRb reaction in the presence of 1064 nm light from the ODT that confines the reactants. We assumed that there is a steady-state population of K$_2$Rb$_2^*$ within the trap that results from a balance between their formation from reactants, their dissociation into products, and their possible photoexcitation (Fig. \ref{figLongLivedComplex}a); and we were interested in how this population depend on the intensity of the ODT light. To this end, simply changing the overall ODT intensity will modify the density and temperature of the KRb sample, thereby confounding our measurements. Instead, we apply a time-modulated potential with two instantaneous intensity levels, $I$ and $I'$, such that the time-averaged ODT intensity remains constant as we vary these levels (Fig. \ref{figLongLivedComplex}b inset). By monitoring the strengths of the K$_2$Rb$_2^+$ signal (resulting from the ionization of K$_2$Rb$_2^*$) during the $I$-phase, we found that both the complex population decreases monotonically with increasing values of $I$, indicating that the complexes are indeed photoexcited by the ODT (Fig. \ref{figLongLivedComplex}b). From a fit to this data, we determined that at an optical intensity of 11 kW/cm$^2$
--  a typical value used to confine ultracold samples -- the complex is almost ten times as likely to undergo photoexcitation as it is to proceed into products.

The ability for K$_2$Rb$_2^*$ to rapidly scatter photons comes as a direct consequence of its prolonged lifetime, and stands in stark contrast to the well-known insensitivity of typical (short-lived) complexes to external influences. This property allows us to efficiently steer the pathway for the ultracold reaction between that occurring on the ground surface and that on an excited surface using light, which we exploited to measure the complex lifetime. By suddenly switching off the ODT light while reactions occur, we quench the complex loss rate to a lower value, resulting in a growth of its equilibrium population (Fig. \ref{figLongLivedComplex}c). From the timescale of this growth, we extracted a lifetime of 360±30 ns, in reasonable agreement with that estimated using the RRKM theory, which are 170±60 ns and 228 ns from our work and the separate work of Yang \textit{et al.} \cite{yang2020global}, respectively. The discrepancy between the two RRKM lifetimes stems mostly from differences in the independently constructed \textit{ab initio} PESs used for calculating the complex's DOS. Using the measured lifetime and the ratio between the rates of complex excitation and dissociation, we determined the rate coefficient for the complex excitation to be $0.42 \pm 0.09 ~\mu\textrm{s}^{-1}/(\textrm{kW/cm}^2)$, in agreement with our theoretical prediction of $0.4^{+0.4}_{-0.2} ~\mu\textrm{s}^{-1}/(\textrm{kW/cm}^2)$.

In a parallel experimental effort, Gregory \textit{et al.} investigated the collisions between ultracold RbCs molecules in the presence of infra-red laser light. Since the reaction RbCs + RbCs $\rightarrow$ Rb$_2$ + Cs$_2$ is endothermic by 29 cm$^{-1}$, and the reactants are prepared in their absolute rovibrational and hyperfine ground states, the Rb$_2$Cs$_2^*$ complex, once formed and absent any external influence, can only dissociate back to a pair of RbCs molecules in their original states, a process termed ``revival'' (Fig. \ref{figLongLivedComplex}c). This is in contrast to exothermic reactions such as KRb + KRb, where the complex is overwhelmingly more likely to dissociate into products than reactants due to much larger number of exit channels associated with the piror. Despite the lack of direct signal on the complex itself, evidences of its photoexcitation were found through how the light influences its revival back into molecules. To this end, the ultracold RbCs sample is confined in an 1064 nm ODT with a time-modulated intensity (Fig. \ref{figLongLivedComplex}e inset), allowing some Rb$_2$Cs$_2^*$ complexes to ``revive'' back into RbCs during the dark periods of the modulation. In the presence of an additional 1550 nm laser beam superimposed on the sample, however, a suppression of the revival was observed, indicating photoinduced complex loss. Specifically, the number of RbCs molecules remaining at the end of a fixed wait was found to decrease as the length of the dark period is reduced (Fig. \ref{figLongLivedComplex}e) or as the intensity of the 1550 nm light is increased. From these measurements, Gregory \textit{et al.} inferred a photo-excitation rate constant of $3^{+4}_{-2} ~\mu\textrm{s}^{-1}/(\textrm{kW/cm}^2)$, which is within an order of magnitude of that measured for K$_2$Rb$_2^*$. They further determined the natural lifetime of the Rb$_2$Cs$_2^*$ complex to be $530\pm60~\mu$s, which is 3 orders of magnitude longer than that of K$_2$Rb$_2^*$. The longer lifetime can be directly correlated to the fact that the number of exit channels in the KRb case, which is $N_0 \sim 700$, is about 3 orders of magnitude higher than that for the RbCs case, which is $N_0 = 1$, while the densities of states of the two systems are similar to within a factor of 2.

The photoexcitation of long-lived complexes by the trapping light discovered in both the KRb + KRb system and the RbCs + RbCs system strongly supports the loss mechanism proposed by Christianen \textit{et al.} for ultracold bilakli molecules \cite{christianen2019photoinduced}. Furthermore, it is now clear that the loss of KRb molecules, previously attributed to chemical reactions on the ground-state pathway \cite{ospelkaus2010quantum}, is in fact due mainly to the photoexcitation of the complex by the ODT at the typical intensity at which it is operated. Based on the similarities in electronic structures and two-body reaction energetics across different bialkali species \cite{byrd2012structure,christianen2019six}, it is tempting to assume that the formation of long-lived complexes and their photoexcitation also explain the losses observed in other ultracold molecular gases such as NaK and NaRb. Such a generalization, however, is complicated by the results from two recent studies on these species. Gersema \textit{et al.} \cite{gersema2021probing} measured the two-body decay rates for ultracold samples of bosonic $^{23}$Na$^{39}$K and $^{23}$Na$^{87}$Rb molecules, respectively, in time-modulated ODTs, and found no evidences of revival. Furthermore, they found the decay rates to be near-universal regardless of the intensity of an additional 1064 nm light superimposed on the sample. Bause \textit{et al.} \cite{bause2021collisions} confined fermionic $^{23}$Na$^{40}$K molecules in a blue-detuned ODT, where the light intensity is three orders of magnitude lower than a conventional red-detuned ODT; yet they still measured losses at near the universal rate limit. These new observations suggests that the complexes involved in certain ultracold bilaklai reactions may possess lifetimes that are orders of magnitude longer than that predicted by the RRKM theory, or that the rates of photoexcitations are orders of magnitude higher than theoretical estimates. On a related note, Nichols \textit{et al.} recently studied the endothermic KRb + Rb collision at ultralow temperatures. By directly probing the transient KRb$_2^*$ complex using the apparatus depicted in Fig. \ref{figMachine}, they determined its lifetime to be 0.39$\pm$0.06 ms, which is five order of magnitude longer than the RRKM estimate ($\sim 1$ ns). The highly non-RRKM behaviors observed across different systems call for further scrutiny into their short-range dynamics.

\subsection{Quantum state distribution of products} \label{subsection: quantum state distribution}

%\sout{In molecular beam experiments, information on how a reaction branches into various available product quantum states has proven indispensable to unraveling the short-range dynamics \cite{yang2007state}. As such,} 
Product state detection of ultracold reactions has long been recognized as an important milestone of the field \cite{dulieu2011physics}. The first examples of such detections can be found in studies involving weakly-bound molecules. Rui \textit{et al.} \cite{rui2017controlled} and Hoffmann \textit{et al.} \cite{hoffmann2018reaction} respectively investigated the reaction between a Feshbach molecule and an atom, and that between two Feshbach molecules. The extremely weakly-bound, translationally ultracold products were detected in a state-resolved fashion through absorption imaging (section \ref{section: production}). Due to the weak binding within Feshbach molecules, the dynamics of these reactions are effectively described by spin-spin interactions rather Coulomb interactions. Wolf et al.\cite{wolf2017state} studied the three-body recombinations of ultracold Rb atoms which yield a weakly bound dimer (Rb$_2$) and an atom. The product state distribution was obtained by a state-selective ionization of the dimers, followed by a measurement of the losses of ultracold Rb atoms incurred by the resulting ions.

\begin{figure}[t!]
\centering
\includegraphics[width=1.0\textwidth]{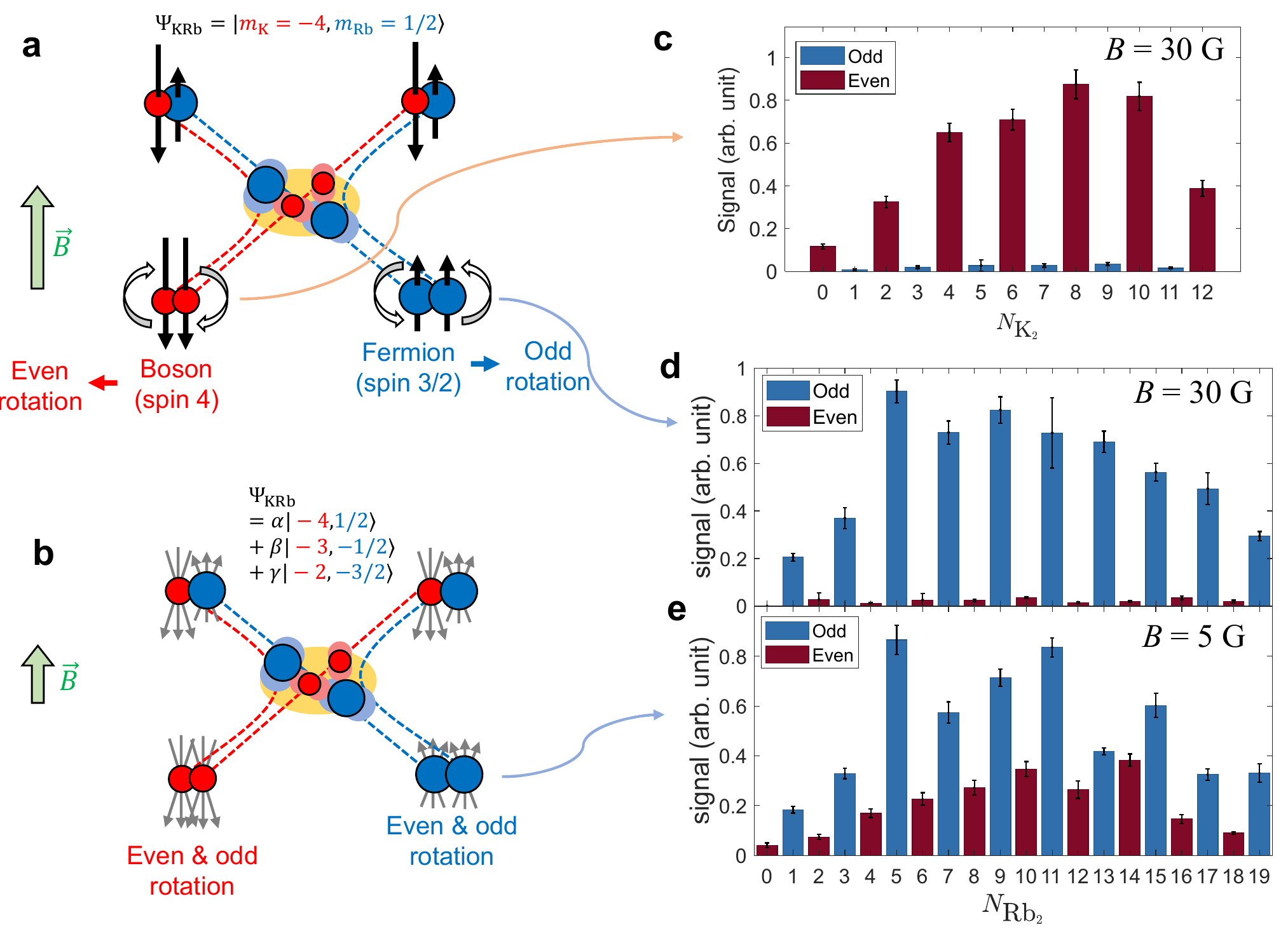}
    \caption{Manipulating the product state distribution of an ultracold reaction via conserved nuclear spins. (a) At a magnetic field of 30 G, the reactant KRb molecules are prepared in a nuclear spin state characterized by a single Zeeman sublevel. The spins and spin projections of the $^{40}$K and $^{87}$Rb nuclei are directly inherited by those of the products, resulting in symmetric spin states. Due to the bosonic (fermionic) statistics of identical K (Rb) nuclei, this results in only even (odd) rotational states for K$_2$ (Rb$_2$), as shown by the state distributions in (c) and (d). (b) Lowering the magnetic field (from 30 G to 5 G) introduces other Zeeman sublevels into the spin state of the reactants, resulting in a superposition of symmetric and antisymmetric spin states in the products. This allows both even and odd rotational states to be populated, as shown by the distribution for Rb$_2$ in (e). $N$ represents the rotational quantum number. Figure adapted from Ref. \cite{hu2020nuclear} with permission.}
\label{figNuclearSpin}
\end{figure}

\begin{marginnote}[]
\entry{REMPI}{resonance-enhanced multiphoton ionization}
\end{marginnote}

In our apparatus (Fig. \ref{figMachine}), the use of ion spectrometry allowed us to generalize product state detection to, in principle, arbitrary ultracold reactions -- including those between deeply-bound molecules, which are of considerable interest from a dynamics point of view. In a study reported in Ref. \cite{hu2020nuclear}, we used resonance-enhanced multiphoton ionization (REMPI) and ion detection to map out the quantum state distributions of the K$_2$ and Rb$_2$ products that emerge from the ultracold KRb + KRb reaction. Due to the limited reaction exoergicity ($\sim 10~\textrm{cm}^{-1}$), only rotational states within the ground vibrational manifolds of the two products are populated. Strikingly, we measured population predominantly in even rotational states of K$_2$ but odd ones of Rb$_2$ (Fig. \ref{figNuclearSpin}b). This strong parity preference turns out to be a consequence of nuclear spin conservation throughout the ultarcold reaction (Fig. \ref{figNuclearSpin}a). Because the KRb reactants are identically prepared in a hyperfine state $| m_{\rm{K}}, m_{\rm{Rb}}\rangle$, where the spins of the constituent K and Rb nuclei have well-defined projections onto a quantization magnetic field (30 G), an invariance of nuclear spins over the reaction results in the product spin states $| m_{\rm{K}}, m_{\rm{K}}\rangle$ and $| m_{\rm{Rb}}, m_{\rm{Rb}} \rangle$, both of which are symmetric under particle exchange. This, in turn, restricts K$_2$ and Rb$_2$ to even and odd rotational states, respectively, due to the exchange statistics of identical bosonic or fermionic nuclei that comprise the products. While nuclear spins acting as spectators is a well-established concept in reaction dynamics, observing it in a system with high spin quantum numbers ($I_{\rm{Rb}} = 3/2, I_{\rm{K}} = 4$) and an extremely long-lived complex is a testament to its robustness.

The observed conservation establishes a well-defined relation between the nuclear spin state of the reactants and the rotational state of the products, providing a unique opportunity for state-to-state control in a complex-forming reaction. In the second part of this study, we leveraged the precise state-control over the reactants afforded by the ultralow temperature to influence the product state distribution. By changing the magnitude of the quantization field, we tuned the reactant nuclear spins state from a single Zeeman sublevel into a coherent superposition of three (Fig. \ref{figNuclearSpin}c). This introduces anti-symmetric components into the product spin states, allowing the opposite parity rotational states to be populated as well (Fig. \ref{figNuclearSpin}d). 

\begin{figure}[t!]
\centering
\includegraphics[width=1.2\textwidth]{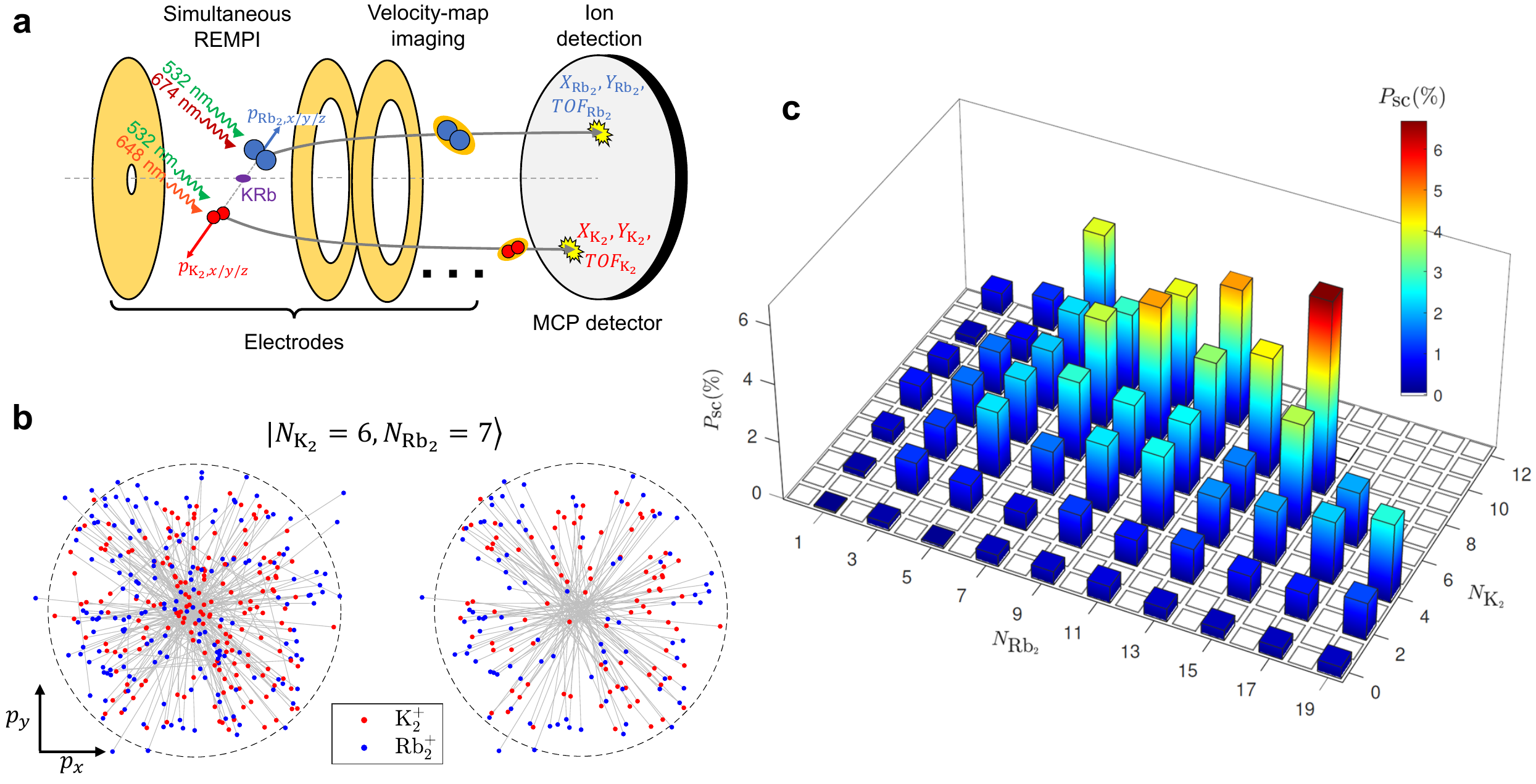}
    \caption{Measuring the pair-correlated product state distribution of the ultracold reaction between KRb molecules. (a) The K$_2$ and Rb$_2$ products emerging from reactions are simultaneously and state-selectively ionized using REMPI. Velocity-map imaging allows the mapping of the three-dimensional momenta of each resulting ion ($p_x,p_y, p_z$) onto the impact location ($X$,$Y$) and time-of-flight ($TOF$) on an MCP detector. (b) Left: two-dimensional momentum image of simultaneously detected K$_2^+$ and Rb$_2^+$ ions associated with the rotational states $N_{\rm{K}_2} = 6$ and $N_{\rm{Rb}_2} = 7$. Each simultaneously detected ion pair is connected by a solid gray line. Right: momentum image of the ions from the left image that additionally satisfy the conservation of linear momentum, $p_{\rm{K}_2,x}+p_{\rm{Rb}_2,x} = 0$, $p_{\rm{K}_2,y}+p_{\rm{Rb}_2,y} = 0$, and $p_{\rm{K}_2,z}+p_{\rm{Rb}_2,z} = 0$. Here, each ion pair corresponds to products from the same reaction event, and we regard it as a coincidence count. (c) The complete pair-correlated product state distribution. The scattering probability into each allowed state-pair is derived from the number of coincidence counts obtained from a measurement similar to that shown in (b). Figure adapted from Ref. \cite{liu2021precision} with permission.}
\label{figCoinDetection}
\end{figure}

While the distributions in Fig. \ref{figNuclearSpin} provide a detailed representation of the reaction's outcome, they remain incomplete due to the absence of correlations between the quantum states of the two products \cite{liu2007product}: for a K$_2$ product in rotational state $| N_{\rm{K}_2} \rangle$, what is the probablity of finding its Rb$_2$ co-porudct in the state $| N_{\rm{Rb}_2} \rangle$? To complete the picture, we measured the populations in pairs of product rotational states, $| N_{\rm{K}_2}, N_{\rm{Rb}_2}\rangle$ (Fig. \ref{figCoinDetection}), of which there are a total of 57 that are allowed by both energy and conservation and parity selection \cite{liu2021precision}. This measurement was powered by a novel state-selective coincidence detection technique, wherein K$_2$ and Rb$_2$ in specific rotational states are simultaneously ionized using REMPI, and their momentum relations are used to determine whether they emerged from the same reaction events (Fig. \ref{figCoinDetection}b).

\begin{figure}[t!]
\centering
\includegraphics[width=1.2\textwidth]{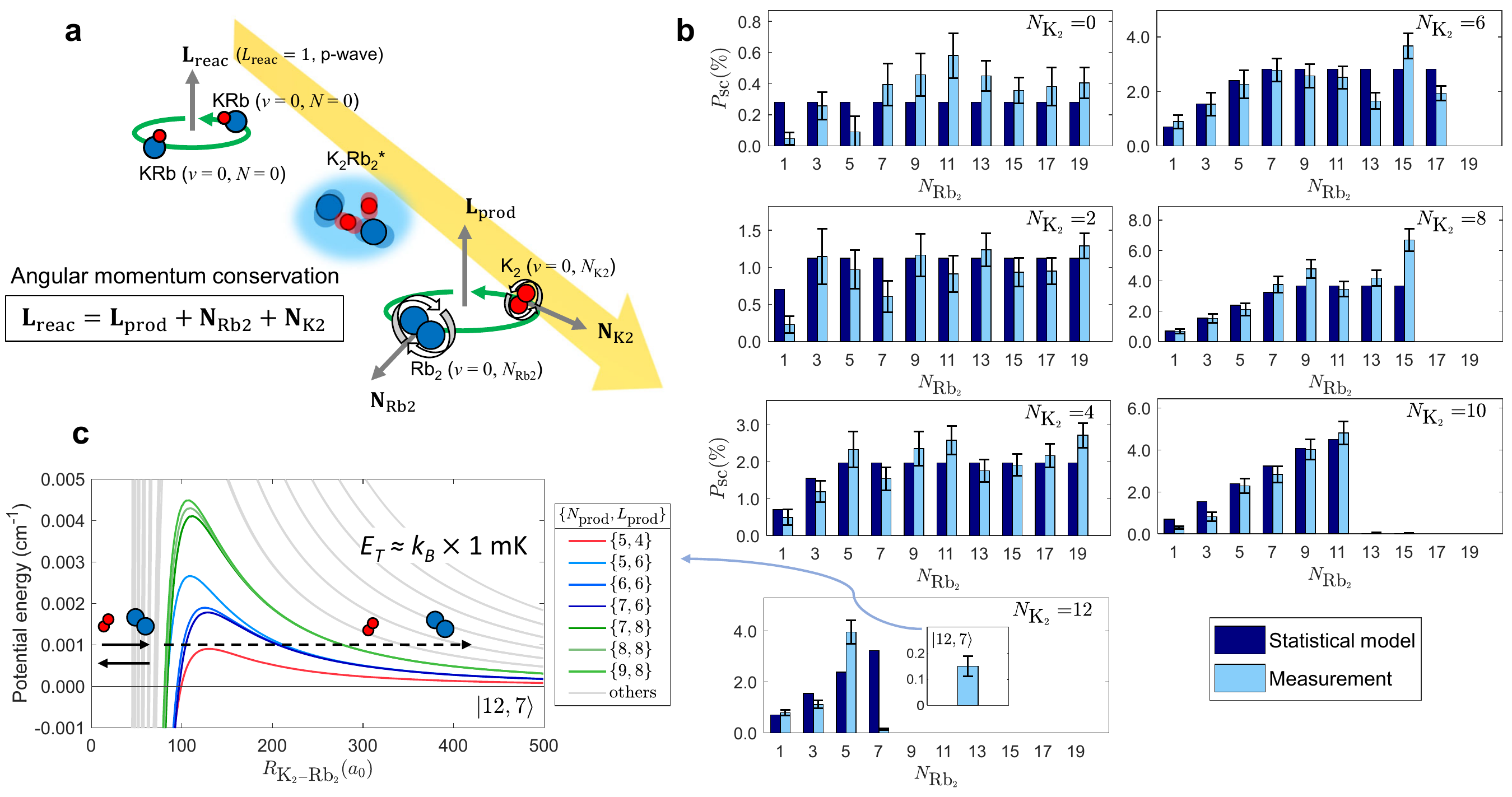}
    \caption{Testing the statistical theory using the product state-distribution of the ultracold KRb + KRb reaction. (a) A schematic illustration of the relations between the various rotational ($\mathbf{N}$) and orbital ($\mathbf{L}$) angular momenta involved in the reaction. At ultrlow temperatures, reactants prepared in their rotational ground states ($N_{\rm{KRb}}$ = 0) react via $p$-wave collisions ($L_{\rm{reac}} = 1$). The rotational angular momenta of the products, $\mathbf{N}_{\rm{K}_2}$ and $\mathbf{N}_{\rm{Rb}_2}$, can take on many different projections with respect to a space-fixed axis, so long as they add with the products' orbital angular momentum ($L_{\rm{prod}} = 1$) to conserve total angular momentum. This results in degenerate scattering channels within each rotational state-pair specified by $| N_{\rm{K}_2}, N_{\rm{Rb}_2}\rangle$. (b) A comparison between the measured product distribution (light blue bars) and that constructed according to the statistical model (dark blue bars). (c) Suppression of product formation into the state-pair $| 12, 7\rangle$ by exit-channel centrifugal barriers. Each potential energy curve corresponds to a specific scattering channel. The tunneling probability calculated based on the degree of suppression indicates that the translational energy for these products is only $\sim 1$ mK. Figure adapted from Ref. \cite{liu2021precision} with permission.}
\label{figProdStateDist}
\end{figure}

The completeness of the above product state measurement, together with the initialization of reactants in a single quantum state, enabled a test of the celebrated statistical theory with unprecedented precision \cite{light1967statistical,pechukas1976statistical}. Statistical theory assumes that a long-lived intermediate complex (such as K$_2$Rb$_2^*$) has sufficient time to ergodically explore the reaction phase space and redistribute its energy among the available modes of motion. This leads to an equal partitioning of scattering probabilities into all allowed product channels, which, for the ultracold KRb + KRb reaction, manifests as the statistical distribution shown in Fig. \ref{figProdStateDist}b. Note that for each state-pair, multiple degenerate scattering channels arise due to the freedom in choosing the relative orientations of the corresponding rotation vectors $\textbf{N}_{\rm{K}_2}$ and $\textbf{N}_{\rm{Rb}_2}$, so long as the total angular momentum of the system is conserved (Fig. \ref{figProdStateDist}a). Comparing the measured distribution to the statistical model, we found good overall agreements but also significant differences for several state-pairs. In particular, the ultralow collision energy of the system allowed us to observed a strong suppression of product formation in the near-threshold state-pair $| N_{\rm{K}_2} = 12, N_{\rm{Rb}_2} = 7\rangle$, due to the presence of exit channel centrifugal barriers (Fig. \ref{figProdStateDist}c). Based on the degree of this suppression, we determined that the products in $| 12, 7 \rangle$ emerge with just $\sim 1$ mK of translational energy, and their orbital motion is dominated by a single $g$ partial wave, which corresponds to $L_{\rm{prod}} = 4$. These ``ultracold'' products provide an unique opportunity for state-to-state control. For example, electric fields may be applied during reactions to slightly modify the reaction exothermicity \cite{meyer2010product,gonzalez2014statistical}, leading to significant changes in the rate of product formation into this state-pair.

The deviations from statistical theory observed in other state-pairs (\textit{e.g.} $| 8, 15 \rangle$, $| 12, 5 \rangle$), on the other hand, cannot be explained by long-range effects, and may be manifestations of short-range dynamics. A definitive explanation of these differences, however, will likely require exact quantum calculations beyond the current state-of-the-art. For the KRb + KRb system, electronic structure calculation is encumbered by the large number of electrons associated with the heavy alkali atoms involved, and scattering calculations for the reaction dynamics is made difficult by the high density-of-state associated with the long-lived complex. In fact, similar computational difficulties is expected for most experimentally accessible ultracold reactions, since the reactants tend to consist of heavy atoms, and their reactions are predominantly complex-forming. While full-dimensional \textit{ab initio} PESs have recently become available for several light bialkali systems \cite{christianen2019six,yang2020global}, tracking the exact reaction dynamics on these surfaces remains challenging \cite{li2020advances}. Yet, the fact that ultracold reactions can be probed on a truly state-to-state level provides an opportunity to rigorously benchmark any future calculations, and should strongly motivate further developments in theoretical methods as well as compulational powers.

While exact calculations for four-atom ultracold reactions are still beyond the horizon, semiclassical methods have been in development by Soley \textit{et al.}\cite{soley2018classical,soley2021reducing} and may be applied to realistic systems in the near future. Meanwhile, exacts calculation of three-atom ultracold reactions have been realized by several groups. Croft \textit{et al.}\cite{croft2017long} performed PES and scattering calculations for the reaction K + KRb $\rightarrow$ K$_2$ + Rb in full dimensionality. The resulting rotationally resolved scattering probabilities were found to follow a Poisson distribution, which was recognized as a signature of chaotic short-range dynamics within the K$_2$Rb$^*$ complex. In a recent study, Kendrick \textit{et al.}\cite{kendrick2021quantum} carried out exact quantum dynamics calculations for the Li + LiNa $\rightarrow$ Li$_2$ + Na reaction, and also observed a Poisson distribution for the state-resolved scattering rates. Furthermore, the inclusion of the first excited PES in this calculation enabled the observation of quantum interference between two pathways around a conical intersection, which results from the different geometric geometric phases incurred along the trajectories\cite{huang2021mechanistic}. In fact, several works by Kendrick and coworkers \cite{kendrick2015geometric,kendrick2018non} have shown interference and geometric phase to be a common theme throughout different ultracold reactions, owing in part to a effective quantization of scattering phase shifts that occur at ultralow collision energies. The intriguing predictions from these theoretical studies provide strong motivations for product-state-resolved experiments on these ultracold systems and beyond.

\section{Summary and outlook}

Advances in AMO techniques over the past decade have enabled the creations of samples of molecules at temperatures as low as a millionth of a degree above absolute zero.
%\sout{in which collisions occur via the single lowest allowed partial wave, and the quantum states are precisely defined for all molecular degrees of freedom.}
Investigations of chemical reactions between these extremely cold and highly controlled molecules revealed behaviors that are markedly different from those that occur under thermal conditions. The majority of ultracold chemistry studies to date focused on the long-range portion of the reaction surface. This now substantial body of work established the extremely sensitivity of ultracold reactivity to details of the long-range forces, which are intimately tied to the quantum states of the reactants. Exploiting this unique feature, many have demonstrated control over the overall rates of ultracold reactions across large dynamic ranges by manipulating reactant states with external fields.

Over the last couple of years the exploration of ultracold chemistry was extended into the short-range thanks to an integration of AMO techniques for preparing the ultracold samples and physical chemistry techniques for probing the ensuing reactions. The ability to directly detect and interrogate the reaction products and intermediates provided a wealth of new information on the reaction dynamics, far beyond what could be achieved from detecting the loss of reactants. In this review, we discussed a series of studies on the short-range dynamics of the ultracold reaction $\rm{KRb} + \rm{KRb} \rightarrow \rm{K}_2\rm{Rb}_2^* \rightarrow \rm{K}_2 + \rm{Rb}_2 $, including the observation of an extremely long-lived intermediate complex, the demonstration of product rotational state control via conserved nuclear spins, and a rigorous test of whether the product state distribution obeys the statistical model following a long-lived complex.

Long-lived complexes appears to be a universal feature in ultracold reactions, and we envision it to be a major topic of research in ultracold chemistry for years to come. While the RRKM theory has correctly predicted the complex lifetimes for some systems, recent studies of several different ultracold reactions all suggest complex lifetimes that are orders of magnitude longer than RRKM predictions. Resolving these apparent discrepancies will surely improve our understanding of reaction dynamics in this new temperature regime. Furthermore, the long complex lifetime, coupled with the observed photo-excitation pathways, opens up the possibility to perform high resolution spectroscopy on the complex, thus elucidating its rovibrational and electronic structures.

At ultralow temperatures, the combined abilities to fully define the quantum states of the reactants and to read out those of the products is a powerful resource for understanding and controlling chemistry. In the KRb + KRb reaction, the realization of product rotational state control via conserved nulcear spins provides a first example of harnessing the unparalleled controllability of ultracold reactants to influence the outcome of the reaction. To this end, an exciting new direction is the coherent quantum control of bimolecular reactions, a concept that is previously demonstrated only on photochemical processes \cite{ohmori2009wave}. As recently proposed by Devolder \textit{et al.}, under ultracold conditions, product channels can be effectively turned on and off by preparing the reactants in different coherent superpositions of initial states \cite{devolder2021complete}. 
%The experimental realization of such a feat will represent a major leap forward in our ability to control reaction dynamics.

Finally, while bialkalis have so far being the dominant subjects of research in ultracold chemistry, innovations in cooling and trapping techniques promise to bring a diverse range of molecular species into the ultracold regime in the near future \cite{wu2017cryofuge}. Recent observations of sub-kelvin collisions between O$_2$ and O$_2$ \cite{segev2019collisions} and between C and O$_2$ \cite{karpov2020low} in a superconducting magnetic trap represent important first steps in this direction.

%Disclosure
\section*{DISCLOSURE STATEMENT}
The authors are not aware of any affiliations, memberships, funding, or financial holdings that
might be perceived as affecting the objectivity of this review. 

% Acknowledgements
\section*{ACKNOWLEDGMENTS}
This work is supported by the U.S. Department of Energy, Office of Science, Basic Energy Sciences, under Award \#DE-SC0019020.

% References
%
% Margin notes within bibliography
% \section*{LITERATURE\ CITED}

\bibliography{refs.bib}
\bibliographystyle{ar-style3}

\end{document}